\documentclass[%
 aip,
 amsmath,amssymb,
 reprint,%
tightenlines
]{revtex4-2}

\usepackage{graphicx}
\usepackage{dcolumn}
\usepackage{bm}

\usepackage[utf8]{inputenc}
\usepackage[T1]{fontenc}
\usepackage{mathptmx}
\usepackage{etoolbox}
\usepackage{xspace}
\usepackage{siunitx}
\usepackage{comment}

\usepackage{titlesec}
\titlespacing*{\section}
{0pt}{3.5ex plus 1ex minus .2ex}{2.3ex plus .2ex}
\titlespacing*{\subsection}
{0pt}{3.5ex plus 1ex minus .2ex}{2.3ex plus .2ex}

\newcommand{\ExB}{$\mathrm{E\times B}$\xspace}
\newcommand{\degc}{$^{\circ}\mathrm{C}$\xspace}
\newcommand{\eref}[1]{Eq.~(\ref{#1})}
\newcommand{\Eref}[1]{Equation~(\ref{#1})}
\newcommand{\fref}[1]{Fig.~\ref{#1}}
\newcommand{\Fref}[1]{Figure~\ref{#1}}

\makeatletter
\def\@email#1#2{%
 \endgroup
 \patchcmd{\titleblock@produce}
  {\frontmatter@RRAPformat}
  {\frontmatter@RRAPformat{\produce@RRAP{*#1\href{mailto:#2}{#2}}}\frontmatter@RRAPformat}
  {}{}
}%
\makeatother
\begin{document}

\preprint{AIP/123-QED}

\title[]{Conceptual Design of a Doppler Spectrometer\\for 10$^2$ m/s Cross-Field Flows in Tokamak Divertors}
\author{K. Fujii}
\email{fujiik@ornl.gov}
\affiliation{ 
  Oak Ridge National Laboratory, Oak Ridge, TN 37831, USA
}%

\author{R. Sano}
\author{T. Nakano}
\affiliation{
  National Institutes for Quantum and Radiological Science and Technology, Naka 311-0193 Japan
}%

\author{G. Ronchi}%
\author{J.-S. Park}%
\author{J. Lore}%
\author{M. Shafer}%
\affiliation{ 
  Oak Ridge National Laboratory, Oak Ridge, TN 37831, USA
}%

\author{T.M. Biewer}%
\affiliation{ 
  Oak Ridge National Laboratory, Oak Ridge, TN 37831, USA
}%

\date{\today}

\begin{abstract}
It has been theoretically predicted that the \ExB drift caused by the spontaneously generated potential in scrape-off-layers (SOLs) and divertors in tokamaks is of a similar size to the poloidal component of the parallel flow and turbulent flow, thereby it significantly impacts on the plasma transport there. 
Many experiments indeed have implied the role of the electric potential, however, its direct observation through its \ExB flow measurement has never been realized because the drift velocity ($10^2$--$10^3$ m/s) is significantly below the detection limit of existing diagnostics. 
To realize a cross-field ion flow measurement, variety of systematic uncertainties of the system must be narrowed down.
Here, we develop a conceptual design of the Doppler spectrometry that enables to measure the impurity flows with $10^2$-m/s accuracy, based on an in-situ wavelength-calibration techniques developed in astrophysics field, the iodine-cell method.
We discuss its properties and applicability.
In particular, the scaling relation of the wavelength accuracy and various spectroscopic parameters is newly presented, which suggests the high importance of the wavelength resolution of the system.
Based on transport simulations for the JT-60SA divertor, the feasibility of the system is assessed.
\end{abstract}

\maketitle

\section{\label{sec:introduction}Introduction}
\begin{figure*}[tbp]
  \includegraphics[width=15cm]{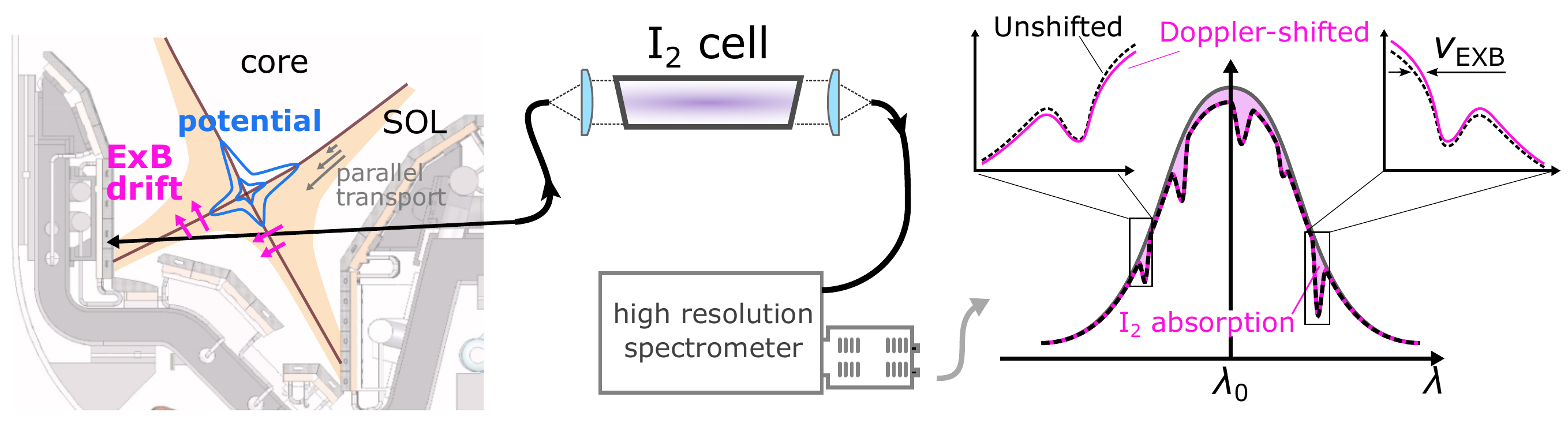}
  \caption{%
      A schematic illustration of the proposed spectroscopic system to measure \ExB flow in tokamak SOLs and divertors.
      (left) Impurity emission will be observed on the poloidal plane. 
      Here, a potential hill in the X-point for a forward-B configuration tokamak is depicted, which causes \ExB flow from the outer divertor to the private flux region. 
      (center) The emission will pass through an iodine-gas cell and then be introduced into a high-resolution spectrometer.
      (right) A spectrum expected to be observed. Absorption lines by iodine molecules will be embedded in the spectrum, which give an absolute wavelength ruler.
      Based on the \textit{in situ} calibration, a tiny Doppler shift by the \ExB flow ($\approx 10^{2}$ m/s, $10^{3}$ times smaller than the thermal width) will be extracted.
  }
  \label{fig:summary}
\end{figure*}

The low-temperature, open-field-line scrape-off-layers (SOL) and divertors are critical regions in a tokamak reactor, acting as the interface between the hot core volume and the plasma facing components (PFCs). 
Extreme particle and heat transport in the SOL can lead to melting and erosion of the PFCs. 
%
One of the important but still experimentally inaccessible parameters for the SOL physics is the electric potential. 
It has been theoretically pointed out that the \ExB drift caused by the spontaneously generated potential (illustrated in \fref{fig:summary}) is of a similar size to the poloidal components of the parallel flow and turbulent cross-field flow, and thereby significantly impacts the in/out and up/down asymmetry of the divertor flux, the impurity transport and retention, and the attach/detach transitions in divertors~\cite{Rozhansky2018-vh,Wensing2020-jp}.
Although experiments have suggested the important roles of the electric potential~\cite{Jaervinen2020-bz}, there are still inconsistencies in the literature, e.g., the in/out balance of the particle and energy fluxes~\cite{Chankin1997-nj}. 

The electric field has been almost always the most inaccessible parameter in plasma physics. 
One simple way to obtain the electric field in magnetized plasmas is to measure the plasma flow due to \ExB drift $v_{E\times B}$, from which the field strength can be deduced from $v_{E\times B}=(\mathbf{E}\times \mathbf{B})/B^2$. 
Indeed, this technique has been successful to evaluate the electric field strength inside the separatrix and has provided the first experimental evidence to the theory of H-mode~\cite{Groebner1990-ar,Ida1990-mu}, which has predicted the build up of the electric field in the separatrix~\cite{Itoh1989-oa}.
The use of the \ExB flow measurement method may also provide the field strength in the SOLs, but the challenge is the much weaker field strength there, $\approx \mathrm{10^3\;V/m}$. 
As the magnetic field of a typical machine is $\approx 3$ T, at least $10^2$ m/s resolution is required to measure the field strength.

Various flow-speed measurements have been conducted in magnetized plasmas. 
Charge-exchange spectroscopy has been utilized to measure the \ExB drift speed in the separatrix.
Special attention has been paid to improve the measurement accuracy, and $10^3$ m/s accuracy has been realized, which is sufficient to measure the field strength there $\gtrsim \mathrm{10^4\;m/s}$~\cite{Yoshinuma2010-gc}.
The toroidal flow speed has been also measured from impurity emission lines in the visible and X-ray ranges~\cite{Isler1999-jq, Roquemore1999-mo}. 
The toroidal flow speed is also $\gtrsim \mathrm{10^4\;m/s}$, which is approachable by conventional techniques.
Laser-based techniques, such as laser-induced fluorescence spectroscopy, has demonstrated measurements of slower flows ($\gtrsim 10^2$ m/s), but its application is limited to small-scale experiments, as this method requires long measurement times ($\gtrsim 10^1$ s) and complex measurement layouts.
For these reasons, \ExB flow measurement in the SOLs has not been realized yet.

\section{\label{sec:challenges}Technological Challenges}

The $10^2$-m/s accuracy in ion flow speed corresponds to 
$2\times 10^{-4}$ nm wavelength accuracy for 600 nm light.
The Doppler width of an impurity ion in a typical divertor plasmas is $10^{-1}$ nm (assuming carbon ions with 30 eV temperature), and the instrumental width of a typical high-resolution spectrometer is $10^{-2}$ nm (1-m focal length, 2400 grooves/mm grating).
The Doppler shift to be measured is $\approx 10^{3}$ times smaller than the spectral width and $\approx 10^{2}$ times smaller than the instrumental width.

At this accuracy level, subtle systematic effects that are usually ignored in conventional spectroscopy become important.
For example, 1\degc change in the room temperature causes \SI{10}{\micro\meter} thermal expansion of a spectrometer ($2\times 10^{-5}\;^{\circ}\mathrm{C}^{-1}$ thermal expansion rate of aluminum and 0.5 m focal length, and 2400 grooves/m grating density is assumed), which corresponds to $\approx 10^{-3}$ nm shift for a conventional spectrometer.
The mechanical accuracy of the entrance slit also affects the calibration ($\approx 10$\% of the instrumental width).

Another nonnegligible effect is the illumination pattern of the first optics of a spectrometer. 
Since the magnitude of the optical aberration is bigger at the optics edge than on the optical axis, the angular distribution of the light introduced into a spectrometer affects the instrumental profile, leading to a shift in the wavelength calibration (with the order of $10^{-1}$ of the instrumental width, which is $\approx 10^{-3}$ nm.)
The angular distribution of the illumination on a multi-mode fiber end will be conserved at the other end even with the long travel distance. 

In order to compensate for these effects and their drift, an \textit{in situ} wavelength calibration is required. 
Various methods have been proposed.
Allen \textit{et al} have developed a method using a reference laser light~\cite{Allen2018-br}. 
They have proposed to use an optical parametric oscillator laser, which can tune the laser wavelength over almost the entire visible range. 
They have proposed to calibrate the system before and after every plasma experiment with the laser light.
Although the laser wavelength can be tuned with $10^{-5}$ nm accuracy, the illumination patterns must be matched exactly between the reference light and plasma light.
Thermal drift during an experiment might also be problematic for long experiments, i.e. more than a minute in duration.
Ida \textit{et al} has proposed to measure an emission line by two oppositely directed sight lines~\cite{Yoshinuma2010-gc}. 
By differentiating two spectra, some of the systematic effects can be compensated. 
They have achieved $10^{-3}$ m/s accuracy in charge exchange spectroscopy for the Large Helical Device.

In this work, we propose the use of another innovative method, the iodine-cell method, which has been proposed to compensate for all of the systematic uncertainties described above.

\section{Iodine-cell Method}

A schematic illustration of the principle of the iodine-cell method and its application to divertor spectroscopy is shown in \fref{fig:summary}.
The light from the targeted plasma (star light in the case of astrophysical observations) goes through an iodine cell before being incident on the spectrometer slit.
Iodine vapor has an enormous number of absorption lines in the visible range (particularly 500--600 nm range), which will be embedded in the incident spectrum (\fref{fig:summary}~right).
Since the absorption profile of iodine molecules has been well investigated, these lines can be used as a \textit{wavelength ruler} to carry out an \textit{in situ} wavelength calibration only from the spectrum.
The advantage of this method is that the target light and the reference light (absorption lines) will be measured exactly at the same space and time. 
Therefore all of the uncertainty sources described above are automatically compensated for.
Using this method, Butler \textit{et al} and other groups have achieved better than 10 m/s accuracy in the Doppler shift measurement for astrophysical studies~\cite{Butler1996-ln, Sato2002-ly}.

\subsection{A small-scale experiment}

\begin{figure*}[tbp]
  \includegraphics[width=17cm]{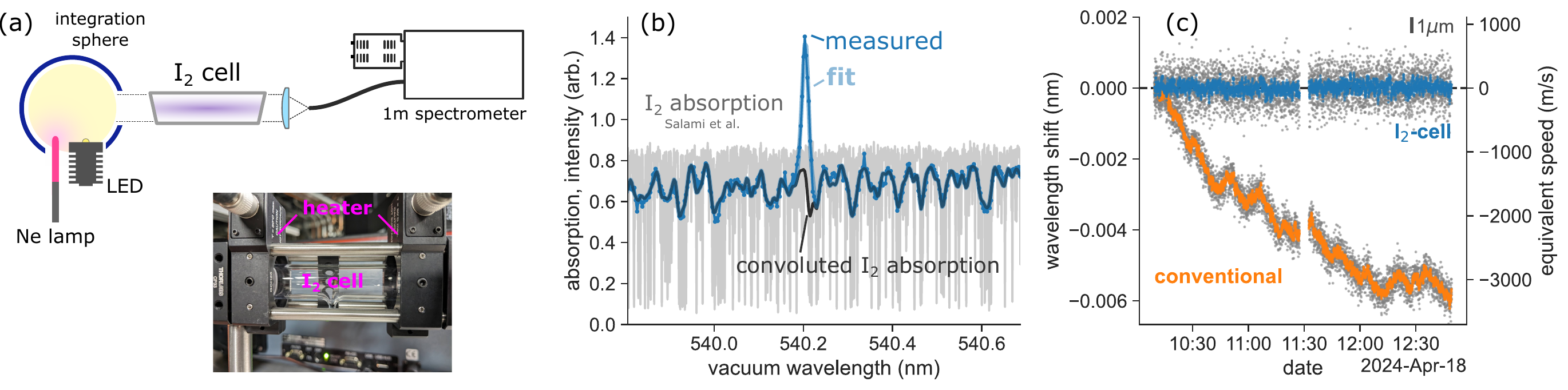}
  \caption{%
      (a) A schematic illustration of the experimental setup for testing the iodine-cell method.
      (b) (markers) The spectrum measured for an LED lamp and neon lamp through the iodine cell. The central peak is Ne I 540.20631 nm line. The dips in the continuum emission are due to the iodine absorption.
      (light gray curve) The iodine absorption spectrum obtained by Salami \textit{et al.}~\cite{Salami2005-na} taken with $4\times 10^{-4}$ nm resolution.
      (black curve) Its convolution with a Gaussian function with a 0.01 nm width.
      (blue bold curve) shows the fit for the absorption and neon emission.
      (c) Temporal evolution of the spectrometer drift over 3 hours. 
      The lower curve shows the measured wavelength of the neon line based on a conventional calibration, where we conducted the calibration only once at the beginning. This drifts exceed more than $5\times 10^{-3}$ nm over the time of observation.
      The 10-min oscillation is probably due to the air-conditioning of the lab space.
      The upper curve shows the measured wavelength of the line based on the iodine-cell method.  
      The scatter is $\approx 10^{-4}$ nm, which is due to the limited SNR of the spectrum (see Sect.\ref{sec:snr}).
      Note that the instrumental resolution of this system is $10\times 10^{-3}$ nm, which is larger than the vertical axis range. 
      The spatial scale on the detector is shown by the scale bar. 
  }
  \label{fig:iodine}
\end{figure*}

\Fref{fig:iodine}~(a) shows a small-scale experimental setup for the iodine-cell method which is built at Oak Ridge National Laboratory by the authors.
The iodine cell (Thorlabs, inc. GC19100-I) is a vacuum quartz cell (100 mm length, 17.5 mm inner diameter) enclosing a tiny iodine solid. 
The gas density in the cell is determined by the saturation pressure of the iodine, which is controlled by a cell heater (Thorlabs, inc. GCH25-75).

We observed a broadband light-emitting-diode (LED) light (Thorlabs, inc. MCWHL7) and a neon lamp emission (Newport corp. Model 6032) fed into an integration sphere (Labsphere, inc., 3-inch diameter).
The light from the integration sphere is introduced into an optical fiber through the iodine cell (heated at 40\degc) and measured its spectrum by a 1-m spectrometer (McPherson model 2051, 1800 /mm grating, 0.01-nm instrumental width with \SI{20}{\micro\meter} slit width). 
The markers in \fref{fig:iodine}~(b) show the measured spectrum. A sharp peak is a neutral neon emission line (NeI 540.20631 nm). Many dips by the iodine absorption can be seen in the continuum emission from an LED. l

The light gray curve in \fref{fig:iodine}~(b) shows the spectrum recorded by Salami \textit{et al.} with $5\times 10^{-4}$ nm resolution~\cite{Salami2005-na}. 
The dark gray curve is the convolution of this spectrum by a Gaussian function with a full-width-half-maximum (FWHM) of 0.01 nm, which mimics the instrumental broadening.
This convolution reproduces well our observation, except for in the vicinity of the neon emission line.

We measure the stability of the iodine-cell calibration method using the neutral neon line.
Here, we assume that the rest wavelength of this neon line is unknown and estimate its rest wavelength from the iodine absorption lines. 
\Fref{fig:iodine}~(c) shows the wavelength drifts observed for this system over three hours. 
The conventional calibration was carried out only once at the beginning, and assumes the same pixel-to-wavelength relation for the rest of the duration.
This calibration drifts more than 0.006 nm during three hours.
As similar to the observation by Allen \textit{et al}.~\cite{Allen2018-br}, a periodic oscillation with about 10-min period is observed, which is probably due to the air-conditioning system of the lab.
On the other hand, the calibration by the iodine-cell method stays the same during the entire time, which indicates the effectiveness of the iodine-cell method.

We note that although in the analysis we fit the measured spectrum by the convoluted iodine absorption spectrum plus a single Gaussian for the neon line, this is not perfectly correct.
Indeed, the complete analysis of this spectrum is challenging, since both the neon line and iodine lines are much narrower than the instrumental width (0.01 nm); the neon line will be absorbed by the iodine first and then convoluted by the instrumental function, and not vice versa. 
This property makes this method suitable to broad emission lines, but is not strictly accurate for lines narrower than the instrumental resolution.

\subsection{Scaling law of the iodine-cell method accuracy\label{sec:snr}}

The \textit{in situ} calibration result in \fref{fig:iodine}~(c) has a scatter of $\approx 10^{-4}$ nm, which is  essentially the uncertainty of the calibration due to the limited SNR of the iodine-embedded spectrum.
Let us consider a scaling law for the calibration uncertainty.

We denote the iodine spectrum convoluted by the instrumental function by $f(\lambda)$.
$f(\lambda)$ is observed at $N$ discrete points $\{\lambda_0, ..., \lambda_i, ..., \lambda_N\}$.
We fit the noisy observation $f(\lambda_i) + \epsilon_i$ by the known function $f(\lambda)$ with an unknown horizontal shift $\Delta_\lambda$, i.e., 
\begin{align}
  f(\lambda_i) + \epsilon_i 
  \approx f(\lambda_i) + \Delta_\lambda \frac{\partial f(\lambda_i)}{\partial \lambda}.
\end{align}
Here, we consider a first-order Taylor expansion.
If $\epsilon_i$ follows a normal distribution with zero mean and standard deviation of $\epsilon$, then the \textit{posterior} distribution of $\Delta_\lambda$ follows a zero-mean normal distribution with the standard deviation of 
\begin{align}
  \label{eq:posterior}
  \epsilon\left[\sum_i^N \left(\frac{\partial f_i}{\partial \lambda}\right)^2\right]^{-1/2},
\end{align}
which is the calibration uncertainty of the iodine-cell method.

The absorption lines of iodine-spectrum distribute almost randomly along wavelength and intensity. 
We observe its convolution with a Gaussian instrumental function (width $\delta_\lambda$, which includes many absorption dips). 
Thus, the convoluted absorption profile $f(\lambda)$ should behave as a Gaussian process with a radial-basis-function kernel having the scaling length $\delta_\lambda$, i.e., 
$K(\lambda, \lambda') = \frac{\alpha}{\delta_\lambda} \exp\left(-\frac{(\lambda - \lambda')^2}{2 \delta_\lambda}\right).$
Here, Gaussian process kernel $K(\lambda, \lambda')$ corresponds to the covariance between $f(\lambda)$ and $f(\lambda')$.
$\alpha / \delta_\lambda$ is the variance of this process, which scales according to $\delta_\lambda^{-1}$, because larger $\delta_\lambda$ averages out more absorption dips and non-absorption peaks.
$\alpha$ is the strength of the absorption, which depends on the wavelength range, cell temperature, and cell length.
As a property of the Gaussian process, $\partial f / \partial \lambda$ also follows a Gaussian process, the variance of which is $\mathbb{E}\left[\sum_i^N (\frac{\partial f_i}{\partial \lambda})^2 / N\right] = \alpha \delta_\lambda^{-3}$. 

Its substitution to \eref{eq:posterior} leads to the following scaling law for the calibration uncertainty ($\sigma_\lambda$),
\begin{align}
  \label{eq:scaling}
  \sigma_\lambda \propto \epsilon N^{-\frac{1}{2}} \delta_\lambda^{\frac{3}{2}}.
\end{align}

\begin{figure}[tbp]
  \includegraphics[width=7cm]{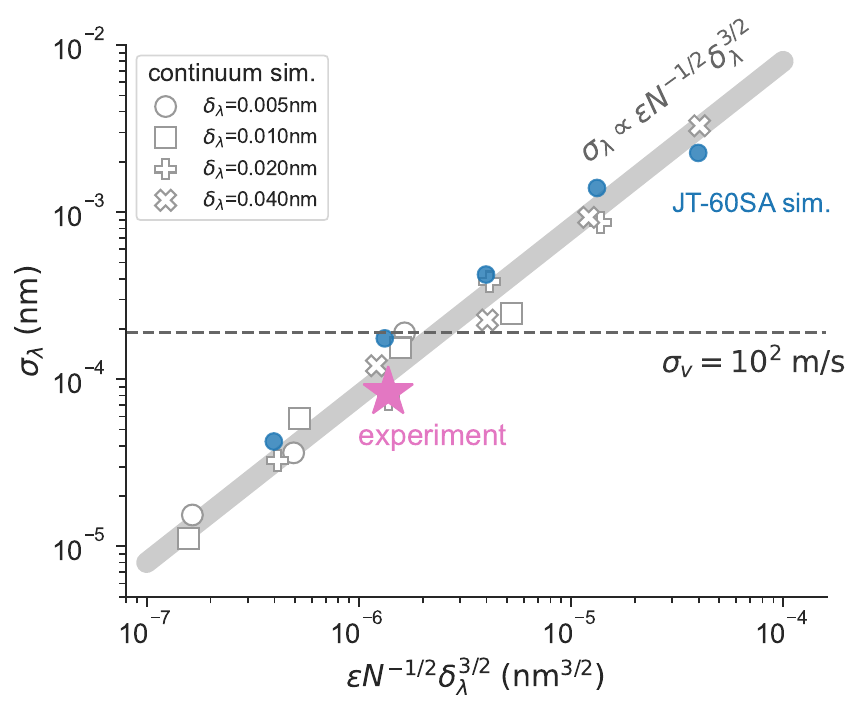}
  \caption{%
      Scaling law for the iodine-cell method accuracy $\sigma_\lambda$,
      as a function of the relative amplitude of the noise $\epsilon$, number of points to be used for the analysis $N$, and the instrumental width of the spectrometer $\delta_\lambda$.
      (Star marker) The accuracy obtained from the experiment, where $\epsilon=0.02, N=400$, and $\delta_\lambda = 0.01$ nm.
      (Open markers) The accuracy obtained from the simulated spectrum of white light and the absorption spectrum. $N=400$, and $\delta_\lambda$ (indicated by marker shapes) and $\epsilon$ are varied.
      (Solid circles) The accuracy obtained for the JT-60SA synthetic diagnostics. $\delta_\lambda = 0.01$ nm. $N \approx 60$ is determined by dispersion of the spectrometer ($1.6\times 10^{-3}$ nm/pixel) and the width of the emission line ($\approx$ 0.1 nm).
  }
  \label{fig:scaling}
\end{figure}

We tested \eref{eq:scaling} by analyzing synthesized iodine spectra.
The original iodine spectrum by Salami \textit{et al}. has $5\times 10^{-4}$ nm resolution (including Doppler broadening) and $1\times 10^{-4}$ nm measurement point spacing~\cite{Salami2005-na}.
We slice the spectrum in the 539.6--540.8 nm range, shift it for various amounts by interpolation to make the $3\times 10^{-3}$ nm interpoint spacing, convolute the instrumental width ($5 \times 10^{-3}$, $1 \times 10^{-2}$, $2 \times 10^{-2}$, and $4\times 10^{-2}$ nm), and add a random Gaussian noise with various amplitude.
By comparing the wavelength shift reconstructed from the synthesized noisy spectrum with the added shift, we estimate the accuracy of the iodine-cell method.
\Fref{fig:scaling} shows the result as a function of the scaled SNR.
The result for different instrumental widths follow the same scaling \eref{eq:scaling}.

\subsection{\label{sec:design}Design consideration of an iodine-cell based spectroscopic system}

\Eref{eq:scaling} helps the design of the Doppler spectrometer for measuring the divertor electric field.
The factor $\epsilon N^{-1/2}$ is proportional to $I^{-1/2}$, where $I$ is the photon number arriving at the detector (assuming Poisson statistics). 
This factor represents the intensity effect: larger number of photons gives better SNR.
The factor $\delta_\lambda^{3/2}$ states the effect of the wavelength resolution of the system.
These two factors are often in a trade-off relation in spectroscopy. 
For example, a wider entrance slit (slit width $w_\mathrm{slit}$) allows more photons, but results in worse spectral resolution.
Since $I\propto w_\mathrm{slit}$ and $\delta_\lambda \propto w_\mathrm{slit}$, we obtain $\sigma_\lambda \propto w_\mathrm{slit}$.
A narrower slit gives higher accuracy for the iodine-cell method.
The longer focal-length ($L$) of the spectrometer gives better resolution ($\delta_\lambda \propto L^{-1}$) but its disadvantage is the reduced photon throughput due to the smaller solid angle of the optics ($I \propto L^{-2}$).
\Eref{eq:scaling} tells $\sigma_\lambda \propto L^{-1/2}$, i.e., a longer focal-length spectrometer still results in higher accuracy. 

$N$ is related also to the wavelength range of the spectrum to be used for the wavelength calibration. 
In divertor spectroscopy, this is limited by the impurity line broadening (if there is no significant continuum light from the plasma). 
The instrumental width should be $\gtrsim 10$ times smaller than the line width.

\section{Synthetic diagnostics for JT-60SA divertor}

\begin{figure}[tbp]
  \includegraphics[width=7cm]{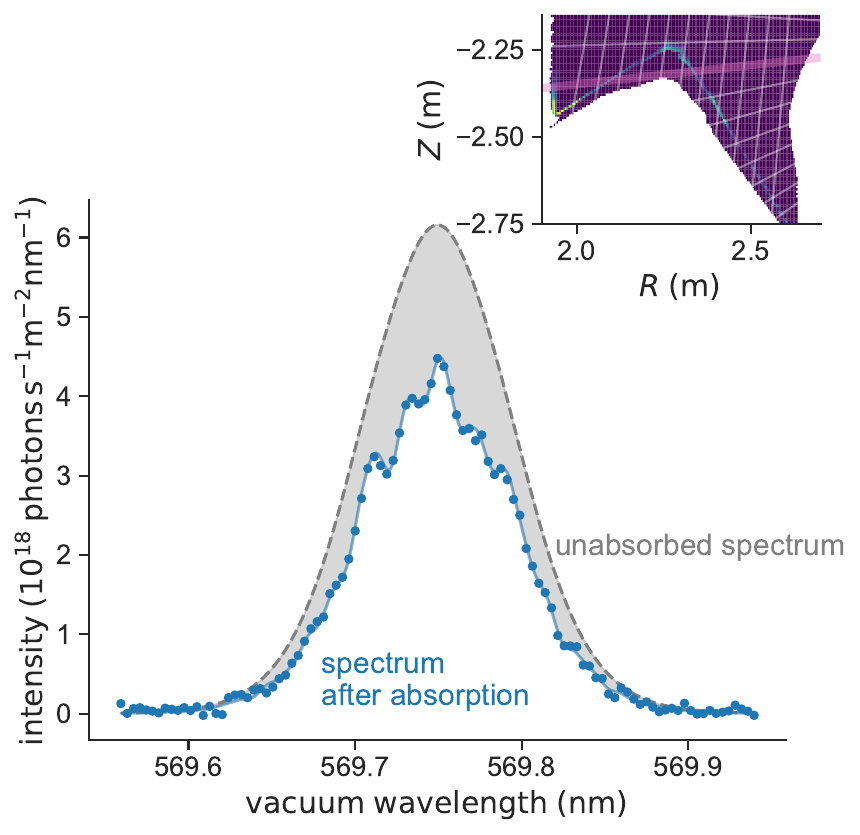}
  \caption{%
      A synthesized spectrum for JT-60SA divertor to be observed. 
      The sight line to be used is shown in the inset (bold line) together with the $\mathrm{C^{2+}}$ emission intensity distribution.
      The spectrum before the absorption is shown by the dashed gray curve, while the spectrum absorbed after an iodine cell is shown by a thin solid blue curve. Markers shows the spectrum with Gaussian noise with $10^2$ SNR.
  }
  \label{fig:jt60}
\end{figure}

The feasibility of the $10^{2}$-m/s Doppler spectrometry is tested based on a synthetic diagnostics for JT-60SA divertor.
SONIC simulation~\cite{Shimizu2009-ta} is conducted to predict the variety of parameters, such as the ion and electron temperature and impurity concentration. 
We assume the following parameters for the SONIC simulation; 
23.0 MW energy flux from core boundary to SOL, $1.2\times 10^{21}$ /s ion flux through the boundary, 3.0\% chemical sputtering yield for carbon, diffusion coefficients for particle ($3.0\mathrm{\;m^2/s}$) and heat ($1.0\mathrm{\;m^2/s}$).
The details of the simulation setup can be found in Ref.~\cite{Yamoto2020-sx}.
\Fref{fig:jt60} inset shows the simulated spatial distribution of $\mathrm{C}^{2+}$ emissivity.
From the spatial distribution of the carbon ion density, electron temperature and density, and ion temperature, as well as the spatial geometry of the sight line (the bold solid line in \fref{fig:jt60} inset) and magnetic field along the sight line, we synthesized the emission spectrum of 569.750-nm line 
($1s^22s3p\mathrm{\;^1P^o}\leftarrow 1s^22s3d\mathrm{\;^1D}$).
The predicted line width is $\approx 0.1$ nm, including the Doppler and Zeeman effect.
Note that since SONIC uses B2 code~\cite{Braams1986-yb} under the hood, but does not yet include an B2.5 extension to calculate electrostatic potential from a continuity equation for the plasma current density~\cite{Braams1996-nv}.
Thus, we use this simulation only to estimate the spectral profile and intensity, but not the shift due to the potential.

We consider to measure the spectrum through an iodine cell (40\degc, 10 cm length) by a conventional high-resolution spectrometer (assuming an equivalence to McPherson model 2051) with 0.01 nm wavelength resolution. 
The wavelength dispersion of the system is assumed to be $1.6\times 10^{-3}$ nm/pixel (\SI{4}{\micro\meter} pixel pitch of the detector is assumed).
The simulated spectrum of the $\mathrm{C}^{2+}$ emission line from JT-60SA divertor before and after the absorption cell are shown in \Fref{fig:jt60} by a dashed curve and markers, respectively. 
Note that we artificially added Gaussian random noise to make the SNR of the spectrum be $10^2$.
Some iodine-absorption dips are apparent in the Doppler-broadened spectrum.

Similar to Sect.\ref{sec:snr}, we add an artificial shift to the divertor spectrum and reconstruct it by using the absorption dip for various values of $\epsilon$.
Although the synthesized spectrum is the sight-line integration of the local emission, which is broadened by the Doppler and Zeeman effects due to the local ion temperature and magnetic field, respectively, in the time of the analysis their detailed spatial distribution is difficult to reconstruct.
Thus, in the analysis we do not consider the Zeeman effect but instead represent the spectrum by a sum of two co-centric Gaussians.
The black circles in \fref{fig:scaling} show the velocity accuracy reconstructed from this analysis, for the various values of SNR ($\epsilon$ = 0.003, 0.01, 0.03, 0.1, and 0.3 from bottom to top). 
Note that the number of pixels $N$ is calculated from the line width and wavelength dispersion of the system, which is $N\approx 60$.
The predicted accuracy is well reproduced by \eref{eq:scaling}.
We found with $\epsilon < 0.01$ the $10^{2}$-m/s accuracy can be achieved.

$\epsilon$ may be estimated from the simulated emissivity, the throughput of the optical system, and the exposure time.
From the simulation, the intensity of this $C^{2+}$ line is $\approx 5\times 10^{17}\mathrm{\;{photons/s\,m^{2}}}$.
Let us assume to measure the emission by a \SI{200}{\micro\meter}-diameter optical fiber with the numerical aperture of 0.2. 
The equivalent area of this fiber image on plasma is $\approx 5\times 10^{-8}\mathrm{\;m^2}$.
We assume the optical efficiency of 0.1 for the light collection and light transfer by the fibers.
The throughput of the spectrometer with \SI{20}{\micro\meter} entrance slit and \#F10 is $\approx 3\times 10^{-3}$. 
To achieve $10^{2}$ SNR for each wavelength channel ($10^{4}\mathrm{\;photons}$), 100 ms exposure time is necessary.

Since a tokamak plasma is often perturbed by edge localized modes, at least 10-ms time resolution is preferred for the quantitative \ExB measurement. 
Thus, we may need $\approx 10$ times better efficiency of the light detection.
An application of a more advanced high-\'etendue spectrometer~\cite{Uzun-Kaymak2012-gn,Fujii2014-ac} will be effective to gain greater throughput. 
For example, a spectrometer consisting of 300-mm focal length lenses, a high-density grism, and the double-pass configuration~\cite{McCubbin1961-yw} will give 10 times greater throughput than McPherson 2051 with keeping the same wavelength resolution.
Also, the use of multiple ($n$) frames for the calibration (but not for the signal) will give us more margin of the measurement accuracy, since the accuracy is increased by a factor of $n^{1/2}$, e.g., assuming $\approx 10^0$-s time scale for the calibration drift will give us 10 times more accuracy.

\section{\label{sec:conclusion}Conclusion}
In this work, we proposed to apply the iodine-cell method to measure $10^2$ m/s flow in divertor regions. 
A scaling law for the calibration accuracy of the iodine-cell method was newly derived, which shows a significant dependence on the wavelength resolution of the spectrometer.  
From synthetic diagnostics for JT-60SA divertor with a $10^{-2}$ nm resolution spectrometer, a $10^{2}$ m/s shift measurement is found to be feasible with 10 ms time resolution, if an advanced spectrometer is employed.

Note that the method proposed here is based on passive spectroscopy, and hence has inherit issues.
The observed emission is sight-line integrated, and there is no spatial resolution along the sight-line. 
In this work, we plan to measure the emission line from $\mathrm{C^{2+}}$, which exists only in limited locations, since they will be ionized quickly in high-temperature plasmas. 
However, in the particular simulation result shown in \fref{fig:jt60}, measurements will be approachable just below the separatrix.
Since the electric field is generated where the temperature gradient is large, and the ions will be ionized in such a region, this limitation is not significant for the electric field measurement.

\section*{Acknowledgements}
This work sponsored by US DOE under DE-AC05-00OR22725.  The work is done under U.S.-Japan Cooperation Programs in Fusion Physics Area (CCFE43-15-FP5-10).

\section*{Data Acknowledgements}
The data that support the findings of this study are available from the corresponding author upon reasonable request.

\bibliographystyle{unsrthtpd}
\bibliography{aipsamp}

\end{document}